\documentclass[aps,prl,twocolumn,showpacs,amsmath,amssymb]{revtex4-1}
\usepackage{amsmath}
\usepackage{graphicx}
\usepackage{subfigure}
\usepackage{epstopdf}
\usepackage{color}
\usepackage{multirow}
\usepackage{setspace}
\usepackage{overpic}
\usepackage{amssymb}
\usepackage[driverfallback=dvipdfmx, bookmarksnumbered, pdfstartview=FitH,colorlinks,urlcolor=blue, citecolor=blue,linkcolor=blue] {hyperref}
\usepackage{lineno}
\usepackage{bm}
\usepackage{rotating}
\usepackage[utf8]{inputenc}

\let\oldequation\equation
\let\oldendequation\endequation

\renewenvironment{equation}
  {\linenomathNonumbers\oldequation}
  {\oldendequation\endlinenomath}

\begin{document}

\title{\bf \boldmath
Absolute Measurements of Branching Fractions of Cabibbo-Suppressed Hadronic $D^{0(+)}$ Decays Involving Multiple Pions
}

\author{
M.~Ablikim$^{1}$, M.~N.~Achasov$^{11,b}$, P.~Adlarson$^{71}$, M.~Albrecht$^{4}$, R.~Aliberti$^{31}$, A.~Amoroso$^{70A,70C}$, M.~R.~An$^{35}$, Q.~An$^{67,53}$, X.~H.~Bai$^{61}$, Y.~Bai$^{52}$
% ...
\\
\vspace{0.2cm}
(BESIII Collaboration)\\
\vspace{0.2cm} {\it
$^{1}$ Institute of High Energy Physics, Beijing 100049, People's Republic of China\\
% ...
$^{4}$ Bochum Ruhr-University, D-44780 Bochum, Germany\\
% ...
$^{11}$ G.I. Budker Institute of Nuclear Physics SB RAS (BINP), Novosibirsk 630090, Russia\\
% ...
$^{31}$ Johannes Gutenberg University of Mainz, Johann-Joachim-Becher-Weg 45, D-55099 Mainz, Germany\\
% ...
$^{35}$ Liaoning Normal University, Dalian 116029, People's Republic of China\\
% ...
$^{52}$ Southeast University, Nanjing 211100, People's Republic of China\\
$^{53}$ State Key Laboratory of Particle Detection and Electronics, Beijing 100049, Hefei 230026, People's Republic of China\\
% ...
$^{61}$ University of Jinan, Jinan 250022, People's Republic of China\\
% ...
$^{67}$ University of Science and Technology of China, Hefei 230026, People's Republic of China\\
% ...
$^{70}$ University of Turin and INFN, (A)University of Turin, I-10125, Turin, Italy; (B)University of Eastern Piedmont, I-15121, Alessandria, Italy; (C)INFN, I-10125, Turin, Italy\\
$^{71}$ Uppsala University, Box 516, SE-75120 Uppsala, Sweden\\
% ...
$^{b}$ Also at the Novosibirsk State University, Novosibirsk, 630090, Russia\\
% ...
}
}

\begin{abstract}
 By analyzing $e^+e^-$ annihilation data with an integrated luminosity of $2.93~\rm fb^{-1}$ collected at the center-of-mass energy $\sqrt s=$ 3.773 GeV with the BESIII detector,
we present the first absolute measurements of the branching fractions of twenty Cabibbo-suppressed hadronic $D^{0(+)}$ decays involving multiple pions.
The largest four branching fractions obtained are
$\mathcal{B}(D^0\to\pi^+\pi^-\pi^0)=   (1.343\pm0.013_{\rm stat}\pm0.016_{\rm syst})\%$,
$\mathcal{B}(D^0\to\pi^+\pi^-2\pi^0)=  (0.998\pm0.019_{\rm stat}\pm0.024_{\rm syst})\%$,
$\mathcal{B}(D^+\to 2\pi^+\pi^-\pi^0)= (1.174\pm0.021_{\rm stat}\pm0.021_{\rm syst})\%$, and
$\mathcal{B}(D^+\to 2\pi^+\pi^-2\pi^0)=(1.074\pm0.040_{\rm stat}\pm0.030_{\rm syst})\%$.
The $CP$ asymmetries for the six decays with highest event yields are also determined.
\end{abstract}

\maketitle

\oddsidemargin  -0.2cm
\evensidemargin -0.2cm

Investigations of hadronic $D^{0(+)}$ decays are of general importance for both charm and bottom physics.
For example, Ref.~\cite{lhcbnote} suggests that hadronic $D^{0(+)}$ decays involving three charged pions are crucial backgrounds for the tests of lepton flavor universality~(LFU) in semileptonic $B$ decays. However, many Cabibbo-suppressed hadronic $D^{0(+)}$ decays with three charged pions are unexplored mainly due to low detection efficiencies and high background contamination.
Precision and comprehensive measurements of the absolute branching fractions (BFs) of these decays provide necessary inputs to unravel the hints of LFU violation observed in semileptonic $B$ decays.

 According to theoretical predictions, the $CP$ violation in charmed hadron decays is expected to be at the order of $10^{-3}$ for singly Cabibbo-suppressed processes, and much
smaller for Cabibbo-favored and doubly Cabibbo-suppressed processes~\cite{ref1,ref2,ref3,ref4,ref5,ref6,ref7,Saur:2020rgd}.
The LHCb experiment reported the first observation of $CP$ violation in $D^0\to K^+K^-$ and $\pi^+\pi^-$ with an asymmetry difference of $\Delta A_{CP}=(1.54\pm0.29)\times 10^{-3}$~\cite{lhcb_D_CP}. A similar magnitude of $CP$ asymmetry is predicted in $D^0\to K^+K^{*-}$ and $D^0\to \rho^+\pi^-$~\cite{ref7}. Refs.~\cite{ref5,ref6} suggest that the $CP$ asymmetries in $D\to \rho \pi$ decays are in the range of $(0.3\sim5)\times 10^{-4}$. Therefore, searching for $CP$ violation in Cabibbo-suppressed $D^{0(+)}$ decays into three pions is an interesting pursuit.

The Cabibbo-suppressed hadronic neutral $D$ decays into $\pi^+\pi^-\pi^0$ and $\pi^+\pi^-\pi^0\pi^0$ also provide a promising way to extract the CKM angle $\gamma$ in $B^+\to D^{(*)0}K^{(*)+}$ due to the similar magnitude of interference amplitudes between $D^0$ and $\bar D^0$ decays into $\pi^+\pi^-\pi^0,\, \pi^+\pi^-\pi^0\pi^0$~\cite{gamma}. More precise measurements of these BFs can improve estimations of the measurement precision for the $CP$ violation phase angle $\gamma$ with these modes.

This Letter reports the first absolute measurements of the BFs of the Cabibbo-suppressed hadronic decays
$D^0\to \pi^+\pi^-\pi^0$,
$\pi^+\pi^-2\pi^0$,
$\pi^+\pi^-2\eta$,
$4\pi^0$,
$3\pi^0\eta$,
$2\pi^+2\pi^-\pi^0$,
$2\pi^+2\pi^-\eta$,
$\pi^+\pi^-3\pi^0$,
$2\pi^+2\pi^-2\pi^0$,  and
$D^+\to 2\pi^+\pi^-$,
$\pi^+2\pi^0$,
$2\pi^+\pi^-\pi^0$,
$\pi^+3\pi^0$,
$3\pi^+2\pi^-$,
$2\pi^+\pi^-2\pi^0$,
$2\pi^+\pi^-\pi^0\eta$,
$\pi^+4\pi^0$,
$\pi^+3\pi^0\eta$,
$3\pi^+2\pi^-\pi^0$,
$2\pi^+\pi^-3\pi^0$ based on 2.93~fb$^{-1}$ $e^+e^-$ annihilation data taken at the center-of-mass energy $\sqrt s=$ 3.773~GeV with the BESIII detector~\cite{3773lumi}. Moreover, $CP$ asymmetries for the six decays with the highest yields are determined.
To date, only the BFs of seven of these decay modes have been measured relative to reference modes and only the $CP$ asymmetries in $D^0(\bar{D}^0)\to \pi^+\pi^-\pi^0$ and $D^\pm\to \pi^+\pi^-\pi^\pm$ have been measured by various experiments~\cite{pdg2020}.
Throughout this Letter, charge-conjugated processes are implied except when discussing $CP$ asymmetries.

The BESIII detector is a magnetic
spectrometer~\cite{BESIII} located at the Beijing Electron
Positron Collider (BEPCII)~\cite{Yu:IPAC2016-TUYA01}. Simulated samples produced with a {\sc geant4}-based~\cite{geant4} Monte Carlo (MC) package including the geometric description of the BESIII detector and the
detector response, are used to determine the detection efficiencies
and to estimate backgrounds. The simulation of $e^+e^-$ annihilations modeled with the generator
{\sc kkmc}~\cite{kkmc} includes the beam-energy spread and initial-state radiation.
The inclusive MC samples consist of the production of $D\bar{D}$
pairs with consideration of quantum coherence for all neutral $D$
modes, the non-$D\bar{D}$ decays of the $\psi(3770)$, the initial-state radiation
production of the $J/\psi$ and $\psi(3686)$ states, and
continuum processes.
Known decay modes are modeled with {\sc
evtgen}~\cite{evtgen} using the BFs taken from
Ref.~\cite{pdg2020}, and the remaining unknown decays
of the charmonium states are modeled with {\sc
lundcharm}~\cite{lundcharm,lundcharm2}. Final-state radiation
from charged final-state particles is incorporated using {\sc
photos}~\cite{photos}.

At $\sqrt s=3.773$~GeV, $D^0\bar D^0$ or $D^+D^-$ pairs are produced without accompanying hadron(s),
thereby offering a clean environment to investigate hadronic $D$ decays with double-tag (DT) method~\cite{Li:2021iwf}.
The single-tag (ST) $\bar D$ candidates are selected by reconstructing a $\bar D^0$ or $D^-$ in the hadronic decay modes:
$\bar D^0 \to K^+\pi^-$, $K^+\pi^-\pi^0$, and $K^+\pi^-\pi^-\pi^+$, and
$D^- \to K^{+}\pi^{-}\pi^{-}$,
$K^0_{S}\pi^{-}$, $K^{+}\pi^{-}\pi^{-}\pi^{0}$, $K^0_{S}\pi^{-}\pi^{0}$, $K^0_{S}\pi^{+}\pi^{-}\pi^{-}$,
and $K^{+}K^{-}\pi^{-}$.
Events in which a signal candidate is reconstructed in the presence of an ST $\bar D$ meson
are referred to as DT events.
The BF of the signal decay is determined by~\cite{bes3-etaetapi}
\begin{equation}
\label{eq:br}
{\mathcal B}_{{\rm sig}} = N_{\rm DT}/(N^{\rm tot}_{\rm ST}\cdot\epsilon_{{\rm sig}}),
\end{equation}
where
$N^{\rm tot}_{\rm ST}=\sum_i N_{{\rm ST}}^i$ and $N_{\rm DT}$
are the total yields of the ST and DT candidates in data, respectively.
The ST yield for the tag mode $i$ is $N_{{\rm ST}}^i$, and
the efficiency $\epsilon_{{\rm sig}}$ for detecting the signal $D$ decay
is averaged over the tag modes $i$.

The selection criteria of $K^\pm$, $\pi^\pm$, $K^0_S$, $\pi^0$, and $\eta$, are the same as those used in the analyses presented in
Refs.~\cite{bes3-etaetapi,bes3-etaX}.
For $\bar D^0\to K^+\pi^-$, the backgrounds from cosmic rays and Bhabha events are rejected by using the same requirements described in Ref.~\cite{deltakpi}.
For $\bar D^0\to K^+\pi^-\pi^-\pi^+$, the $\bar D^0\to K^0_SK^\pm\pi^\mp$ decays are suppressed by requiring the invariant masses of all $\pi^+\pi^-$ pairs to
be outside the mass window $(0.483,0.513)$~GeV/$c^2$.

Tagged $\bar D$ mesons are identified using two variables: the energy difference $\Delta E_{\rm tag} \equiv E_{\rm tag} - E_{\rm b}$
and the beam-constrained mass $M_{\rm BC}^{\rm tag} \equiv \sqrt{E^{2}_{\rm b}-|\vec{p}_{\rm tag}|^{2}}$. Here, $E_{\rm b}$ is the beam energy,
$\vec{p}_{\rm tag}$ and $E_{\rm tag}$ are the momentum and energy of $\bar D$ in the rest frame of $e^+e^-$ system, respectively.
The $\Delta E_{\rm tag}$ of ST $\bar D$ candidates must be in the range $(-55,40)$~MeV for the tag modes involving $\pi^0$
and $(-25,25)$~MeV for the other tag modes, due to differing resolutions.
For each tag mode, if there are multiple candidates in an event,
only the one yielding the smallest $|\Delta E_{\rm tag}|$ is accepted.

To extract the yields of ST $\bar D$ candidates for individual tag modes, binned maximum-likelihood fits are performed to the corresponding $M_{\rm BC}^{\rm tag}$
distributions of the accepted ST candidates following Ref.~\cite{bes3-etaX}.
The $\bar D$ signal is modeled by an MC-simulated shape convolved with
a double-Gaussian function describing the resolution difference between the data and MC simulation.
The combinatorial background shape is described by an ARGUS function~\cite{ARGUS}.
The total yields of the ST $\bar D^0$  and $D^-$ candidates in data are $(232.8\pm0.2_{\rm stat})\times 10^4$ and
$(155.8\pm0.2_{\rm stat})\times 10^4$, respectively.

The signal $D$ decays are selected from the remaining tracks and showers recoiling against the tagged $\bar D$ candidates.
To reject the main backgrounds from Cabibbo-suppressed decays containing $K^0_S$, the $\pi^+\pi^-$ and $\pi^0\pi^0$ combinations are required to not fall in the mass windows $(0.468,0.528)$~GeV/$c^2$ and $(0.428,0.548)$~GeV/$c^2$, respectively. These mass windows correspond to at least $4\sigma$ of resolution.

Signal $D$ mesons are identified using the energy difference $\Delta E_{\rm sig}$ and the beam-constrained mass $M_{\rm BC}^{\rm sig}$, calculated similarly to the ST side.
For each signal mode, if there are multiple candidates in an event, only the one with the minimum $|\Delta E_{\rm sig}|$ is chosen.
Signal decays are required to satisfy the $\Delta E_{\rm sig}$ requirements shown in Table~\ref{tab:DT}.

For each signal decay mode, the signal yield ($N_{\rm DT}$) is obtained from a two-dimensional (2D) unbinned maximum-likelihood
fit~\cite{cleo-2Dfit} to the $M_{\rm BC}^{\rm tag}$ versus $M_{\rm BC}^{\rm sig}$ distribution of the accepted DT candidates.
See Fig. 2 in the Supplemental Material~\cite{Supplemental} for an illustration of the 2D distribution.
The signal events concentrate around $M_{\rm BC}^{\rm tag} = M_{\rm BC}^{\rm sig} = M_{D}$,
where $M_{D}$ is the averaged mass of $D$~\cite{pdg2020}.
The events with correctly reconstructed $D$ ($\bar D$) and incorrectly
reconstructed $\bar D$ ($D$), referred to as BKGI, spread along the lines around
$M_{\rm BC}^{\rm sig}=M_D$ ($M_{\rm BC}^{\rm tag}=M_D$).
Events smeared along the diagonal~(BKGII)
are mainly from the $e^+e^- \to q\bar q$ processes.
Events with uncorrelated and incorrectly reconstructed $D$ and $\bar D$~(BKGIII) disperse across the whole allowed kinematic region.

In the fit, the probability
density functions~(PDFs) for signal, BKGI, BKGII, and BKGIII contributions are constructed as $a(x,y)$,
$b_x(x)\cdot c_y(y;M_{\rm BC}^{\rm end},\xi_{y}) + b_y(y)\cdot c_x(x;M_{\rm BC}^{\rm end},\xi_{x})$,
$c_z(z;\sqrt{2}M_{\rm BC}^{\rm end},\xi_{z}) \cdot g(k;0,\sigma_k)$, and
$c_x(x;M_{\rm BC}^{\rm end},\xi_{x}) \cdot c_y(y;M_{\rm BC}^{\rm end},\xi_{y})$,
respectively.
Here, $x=M_{\rm BC}^{\rm sig}$, $y=M_{\rm BC}^{\rm tag}$, $z=(x+y)/\sqrt{2}$, and $k=(x-y)/\sqrt{2}$.
The PDFs of signal, $a(x,y)$,
$b_x(x)$, and $b_y(y)$, are described by the MC-simulated shapes smeared with individual Gaussian resolution function with parameters derived from the corresponding one-dimensional $M_{\rm BC}$ fits, to consider resolution difference between data and MC simulation. $c_f(f;M_{\rm BC}^{\rm end},\xi_f)$ is an ARGUS function~\cite{ARGUS} ($f$ denotes $x$, $y$, or $z$), where the endpoint $M_{\rm BC}^{\rm end}=1.8865$~GeV/$c^2$ is fixed in the fit.
The Gaussian function $g(k;0,\sigma_k)$ has a mean of zero and a standard deviation parameterized by $\sigma_k=\sigma_0 \cdot(\sqrt{2}M_{\rm BC}^{\rm end}-z)^p$,
where $\sigma_0$ and $p$ are fit parameters. In addition, the yields and shapes of the peaking background (PBKG) components,
which are mainly from $D$ decays into the same final state as a signal mode but involve $K^0_S\to \pi^+\pi^-$ or $\pi^0\pi^0$ and $K^-\to \pi^-\pi^0$ decay, are fixed based on MC simulations and the known BFs of various PBKG components~\cite{pdg2020,Lilanxing}.
Relative to the corresponding signal yields, the PBKG components are 9.5\%, 18.2\%, and 36.2\% for $D^+\to \pi^+4\pi^0$, $D^0\to 2\pi^+2\pi^-\pi^0$, and $D^0\to \pi^+\pi^-3\pi^0$, respectively, and range from 0.1\% to 6.3\% for the other signal decays.

\begin{figure*}[htbp]
  \centering
\includegraphics[width=0.497\linewidth]{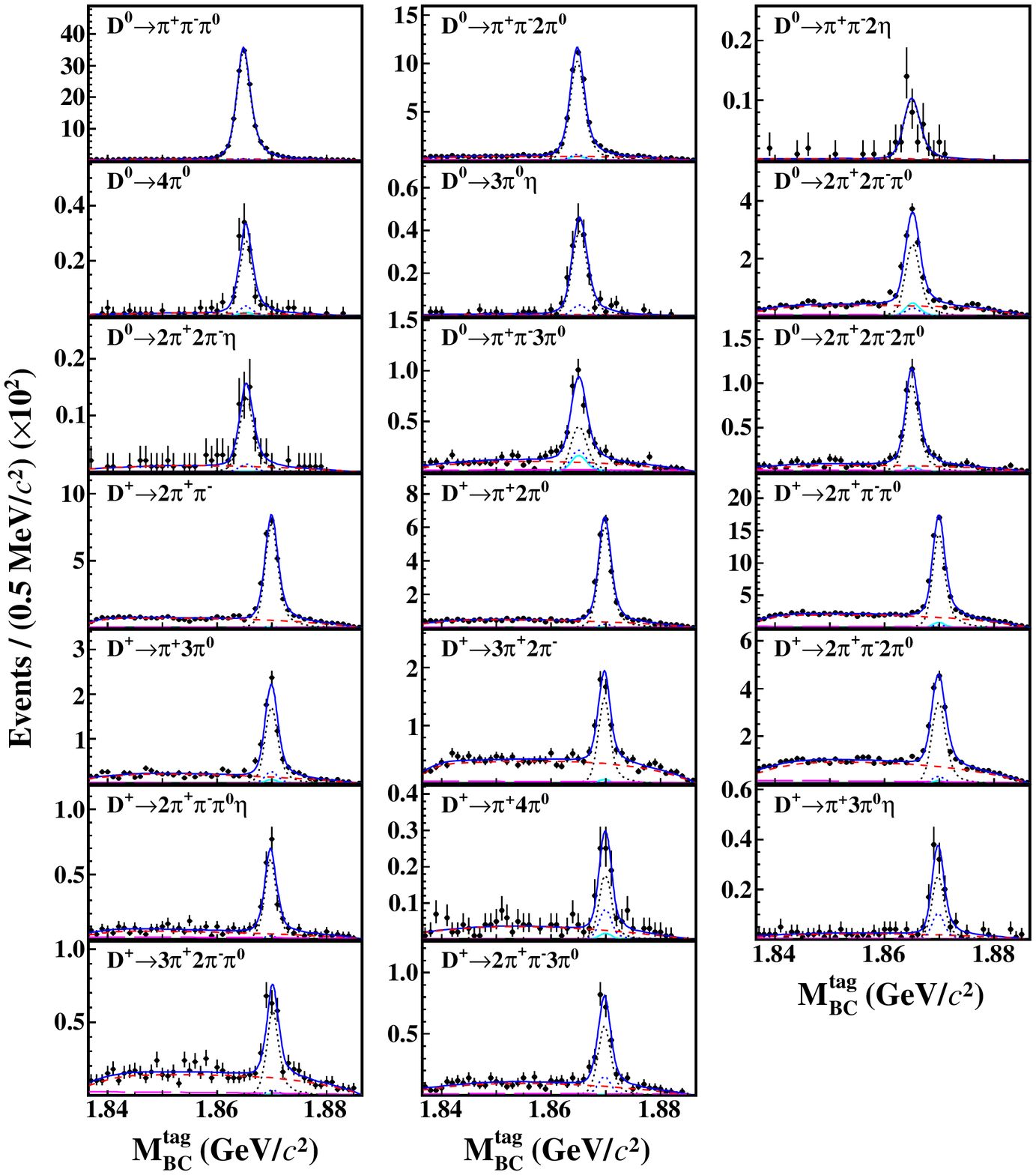}
\includegraphics[width=0.497\linewidth]{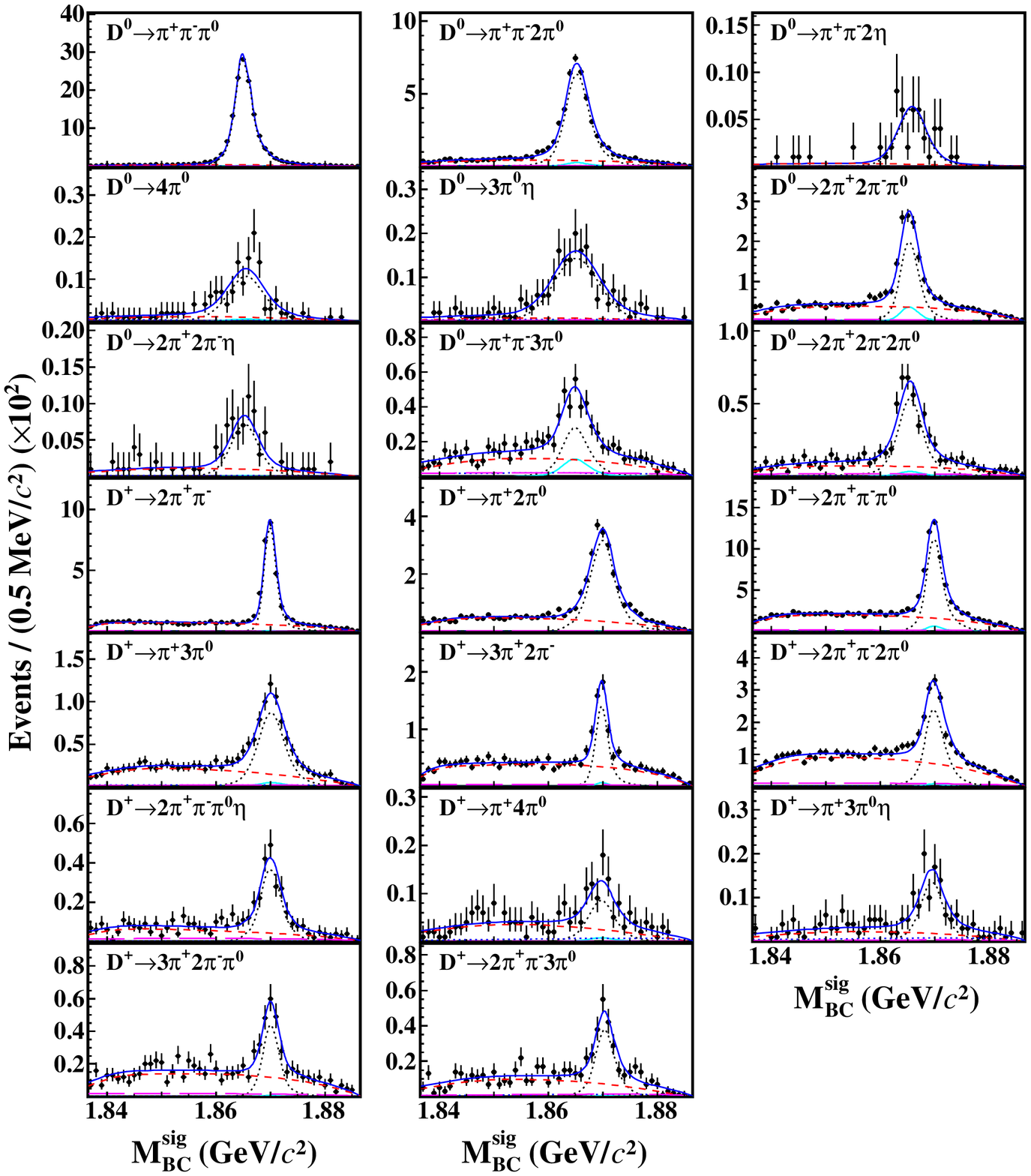}
  \caption{\small
Projections of $M^{\rm tag}_{\rm BC}$ (left) and
$M^{\rm sig}_{\rm BC}$ (right) of the 2D fits to the DT candidate events.
Points with error bars are data and blue solid curves are the fit results.
Black dotted, cyan blue solid, blue dotted, red dotted,  pink long-dashed
correspond to fitted signal, fixed peaking background, BKGI, BKGII, and BKGIII components, respectively.
}
\label{fig:2Dfit}
\end{figure*}

The $M^{\rm tag}_{\rm BC}$ and $M^{\rm sig}_{\rm BC}$ projections of the 2D fits of the DT candidates reconstructed from data are shown in Fig.~\ref{fig:2Dfit}
and the fitted DT yields are summarized in Table~\ref{tab:DT}.

To determine the signal efficiencies ($\epsilon^{}_{{\rm sig}}$), the three-body decays are simulated with a modified data-driven generator BODY3~\cite{evtgen},
which was developed to simulate different intermediate states in data for a given three-body final state.
The Dalitz plot from data, corrected for backgrounds and efficiencies, is taken as input for the BODY3 generator.
The efficiencies across phase space are obtained with MC samples generated according to a phase space distribution.
Each of the four-body and five-body decays is simulated with a mixed signal MC sample.
These decays generated with phase space models including contributions from
$\eta$, $\omega$, $\eta^\prime$, $\rho(770)$, $a_0(980)$, $a_1(1260)$, $b_1(1235)$ and $\phi$
intermediate states are mixed with fractions obtained by examining the corresponding
invariant mass spectra.
The data distributions for momenta and $\cos\theta$ (where $\theta$ is the
polar angle of particle in the $e^+e^-$ rest frame) of the daughter particles, and the invariant masses of each of the two- and multi-body particle combinations agree with the MC simulations. See Figs.~3-11 in the Supplemental Material~\cite{Supplemental} for explicit comparisons.

The results for $N_{{\rm DT}}$, $\epsilon^{}_{{\rm sig}}$, and the extracted BFs
are summarized in Table~\ref{tab:DT}.  The smallest statistical significance
is 6.8 standard deviations for $D^+ \to \pi^+ 4\pi^0$ mode.
The signal efficiencies have been corrected by the data-MC differences in the selection efficiencies of $\pi^\pm$ tracking, particle identification~(PID)
procedures and the reconstruction of $\pi^0$ or $\eta$.

The systematic uncertainties relative to the obtained BFs are discussed below.
In the BF determinations using Eq.~(\ref{eq:br}), all uncertainties from selecting the ST $\bar D$ candidates are cancelled in the ratio. Systematic uncertainties in the total yields of ST $\bar D$ mesons related to the $M^{\rm tag}_{\rm BC}$ fits to the ST $\bar D$ candidates, were previously estimated to be
0.5\% for both neutral and charged $\bar D$~\cite{epjc76,cpc40,bes3-pimuv}.

The tracking and PID efficiencies of $\pi^\pm$ are investigated using other DT $D\bar D$ hadronic events.
The differences of efficiencies between data and MC simulations are weighted by the corresponding $\pi^\pm$ momentum spectra of signal MC events.
The systematic uncertainties due to tracking and PID are assigned to be (0.2-0.4)\% per $\pi^\pm$, based on the residual statistical uncertainties of the measured data-MC differences.

The systematic uncertainty of the $\pi^0$ reconstruction is assigned
as (0.4-0.9)\% per $\pi^0$ from studies of DT $D\bar D$ hadronic decay samples of $D^0\to K^-\pi^+$, $K^-\pi^+\pi^+\pi^-$ versus $\bar D^0\to K^+\pi^-\pi^0$, $K^0_S\pi^0$~\cite{epjc76,cpc40}.
Due to limited $\eta$ statistics, the systematic uncertainty for $\eta$ reconstruction is assigned by referring to that of $\pi^0$.

The systematic uncertainty in the 2D fit to the $M_{\rm BC}^{\rm tag}$ versus $M_{\rm BC}^{\rm sig}$ distribution
is examined by varying the smeared Gaussian function ($\pm 1\sigma$),
the endpoint of the ARGUS function ($\pm0.2$\,MeV/$c^2$),
and the fixed PBKG yields ($\pm 1\sigma$ of the quoted BF).
Adding the changes from these three sources in quadrature yields the corresponding systematic uncertainties.

\begin{table}[htbp]\small
\centering
\caption{\small
Requirements of $\Delta E_{\rm sig}$,
DT yields in data (${N_{{\rm DT}}}$),
detection efficiencies ($\epsilon_{\rm sig}$, including the BFs of $\eta$, and $\pi^0$ as well as correction factors described later),
and the obtained BFs (${\mathcal B}_{\rm sig}$).
The first nine modes are $D^0$ decays and the others are $D^+$ decays.
For ${\mathcal B}_{\rm sig}$, numbers in the first and second parentheses are last two digits
of the statistical and systematic uncertainties, respectively.
For ${ N_{{\rm DT}}}$, uncertainties are statistical only.
}\label{tab:DT}
\begin{ruledtabular}
\resizebox{0.49\textwidth}{!}{
\begin{tabular}{lcrrr}
\multicolumn{1}{c} {Decay}&$\Delta E_{\rm sig}$ & \multicolumn{1}{c}{$N_{\rm DT}$}  & \multicolumn{1}{c} {$\epsilon_{\rm sig}$} & \multicolumn{1}{c} {${\mathcal B}_{\rm sig}$}  \\
& (MeV) &  & \multicolumn{1}{c} {(\%)} & \multicolumn{1}{c} {($\times10^{-4}$)}  \\  \hline
$\pi^+\pi^-\pi^0$ &$(-62,36)$ & $12792.6(120.1)$ & $40.91$ & $134.3(13)(16)$ \\
$\pi^+\pi^-2\pi^0$ &$(-75,37)$ & $3783.7(70.5)$ & $16.29$ & $99.8(19)(24)$ \\
$\pi^+\pi^-2\eta$ &$(-37,29)$ & $42.5(6.7)$ & $2.14$ & $8.5(13)(04)$ \\
$4\pi^0$ &$(-105,41)$ & $96.0(11.5)$ & $5.41$ & $7.6(09)(07)$ \\
$3\pi^0\eta$ &$(-82,40)$ & $155.3(14.7)$ & $2.83$ & $23.6(22)(17)$ \\
$2\pi^+2\pi^-\pi^0$ &$(-52,33)$ & $942.4(40.0)$ & $11.70$ & $34.6(15)(15)$ \\
$2\pi^+2\pi^-\eta$ &$(-36,28)$ & $48.5(7.8)$ & $3.46$ & $6.0(10)(06)$ \\
$\pi^+\pi^-3\pi^0$ &$(-76,39)$ & $182.7(20.9)$ & $5.13$ & $15.3(17)(13)$ \\
$2\pi^+2\pi^-2\pi^0$ &$(-64,36)$ & $350.0(22.9)$ & $3.15$ & $47.7(31)(21)$ \\
$2\pi^+\pi^-$ &$(-30,28)$ & $2614.3(58.0)$ & $50.63$ & $33.1(07)(05)$ \\
$\pi^+2\pi^0$ &$(-96,44)$ & $1968.0(51.7)$ & $27.33$ & $46.2(12)(09)$ \\
$2\pi^+\pi^-\pi^0$ &$(-59,35)$ & $4649.5(83.5)$ & $25.42$ & $117.4(21)(21)$ \\
$\pi^+3\pi^0$ &$(-86,39)$ & $573.7(30.2)$ & $8.83$ & $41.7(22)(13)$ \\
$3\pi^+2\pi^-$ &$(-37,33)$ & $462.1(28.7)$ & $16.26$ & $18.2(11)(10)$ \\
$2\pi^+\pi^-2\pi^0$ &$(-74,39)$ & $1207.1(45.4)$ & $7.21$ & $107.4(40)(30)$ \\
$2\pi^+\pi^-\pi^0\eta$ &$(-51,33)$ & $191.4(15.9)$ & $3.17$ & $38.8(32)(12)$ \\
$\pi^+4\pi^0$ &$(-90,41)$ & $56.7(10.4)$ & $1.87$ & $19.5(36)(23)$ \\
$\pi^+3\pi^0\eta$ &$(-66,37)$ & $79.7(10.9)$ & $1.77$ & $28.9(40)(22)$ \\
$3\pi^+2\pi^-\pi^0$ &$(-49,34)$ & $182.8(17.3)$ & $5.02$ & $23.4(22)(15)$ \\
$2\pi^+\pi^-3\pi^0$ &$(-66,37)$ & $185.9(17.0)$ & $3.49$ & $34.2(31)(16)$ \\
\end{tabular}
}
\end{ruledtabular}
\end{table}

The systematic uncertainty due to the $\Delta E_{\rm sig}$ requirement ranges from (0.1-1.3)\% depending on the signal mode.
They are evaluated from the efficiency differences obtained with and without smearing
the $\Delta E_{\rm sig}$ distributions for signal MC events
with parameters derived from $D^0\to \pi^+\pi^-\pi^0$, $D^0\to \pi^+\pi^-2\pi^0$, $D^+\to \pi^+2\pi^0$, and $D^+\to 2\pi^+\pi^-\pi^0$ to get the data-MC differences.

The systematic uncertainty due to the BODY3 generator is considered by varying the number of bins by $\pm 20\%$ and
the systematic uncertainty in the mixed MC model is assigned by varying the fractions of various components by $\pm 1\sigma$ of the quoted BF, when available.  Unmeasured components
are varied by $\pm 25\%$, beyond which comparisons with observed mass spectra are unsatisfactory.
The largest change of the signal efficiencies, (0.2-5.7)\% for various signal modes, are assigned as the corresponding systematic uncertainties.

The systematic uncertainties due to the mass window applied to reject $K_S^0$ events are crosschecked by examining the changes of the BFs by varying the corresponding boundaries of the window by $\pm5$ MeV/$c^2$.  If the difference of the BF is larger than 1 time the statistical uncertainty on the difference (taking the correlated samples into account), it is assigned as the corresponding systematic uncertainty. Otherwise, it is neglected.

The measurements of the BFs of the neutral $D$ decays are affected by quantum correlation effect~\cite{Wilkinson:2021tby}.
To take this effect into account, the $CP$-even fractions in various decays are needed.
The $D^0\to 4\pi^0$ and $D^0\to 3\pi^0\eta$ final states are both $CP$-even eigenstates.
For $D^0\to \pi^+\pi^-\pi^0$, its $CP$-even fraction has been determined to be $0.973\pm0.017$~\cite{CP-3pi}.
For $D^0\to \pi^+\pi^-2\pi^0$, $D^0\to 2\pi^+2\pi^-\pi^0$, $D^0\to \pi^+\pi^-3\pi^0$, and $D^0\to 2\pi^+2\pi^-2\pi^0$,
the $CP$-even fractions are estimated by the $CP$-even tag $D^0\to K^+K^-$ and the $CP$-odd tag $D^0\to K^0_S\pi^0$.
Using the same method as described in Ref.~\cite{QC-factor}
and the necessary parameters quoted from Refs.~\cite{R-ref1,R-ref2,R-ref3},
we obtain the correction factors to account for the quantum correlation effect on the measured
BFs; the results are summarized in Table 3 of the Supplemental Material~\cite{Supplemental}.
After correcting the signal efficiencies by the individual factors, the residual uncertainties are assigned as systematic uncertainties.

The uncertainties of MC statistics for various signal decays, (0.2-1.3)\%, are considered as a systematic uncertainty.
The uncertainties of the daughter BFs of $\eta\to \gamma\gamma$ and
$\pi^0\to \gamma\gamma$ are 0.51\% and 0.03\%, respectively~\cite{pdg2020}.

Adding all individual effects  for each signal decay  quadratically yields the total systematic
uncertainties to be (1.2-11.9)\% depending on the signal mode.
The detailed systematic uncertainties are
given in Table 5 of the Supplemental Material~\cite{Supplemental}.

For the six decay modes with the highest yields, the BFs of $D$ and $\bar D$ decays,
${\mathcal B}^+_{\rm sig}$ and ${\mathcal B}^-_{\overline{\rm sig}}$, are measured separately.
Their asymmetry is determined by
${{\mathcal A}_{CP}^{\rm sig}}=\frac{{\mathcal B}^+_{\rm sig}-{\mathcal B}^-_{\overline{\rm sig}}}{{\mathcal B}^+_{\rm sig}+{\mathcal B}^-_{\overline{\rm sig}}}$.
The obtained BFs and asymmetries are summarized in Table~\ref{tab:CP}.
We find no statistically significant $CP$ violation.
Several systematic uncertainties cancel in the asymmetry: the tracking and PID of $\pi^+\pi^-$ pairs, $\pi^0$ and $\eta$ reconstruction, daughter BFs, $K^0_S$ rejection windows, MC modeling,
and strong phase of $D^0$ decays.
The other systematic uncertainties are estimated separately as above.

\begin{table}[htp]
\centering
\caption{\small
Charge-separated BFs (${\mathcal B}^+_{\rm sig}$ and ${\mathcal B}^-_{\overline{\rm sig}}$),
and their asymmetries (${{\mathcal A}_{CP}^{\rm sig}}$).
The first and second uncertainties are statistical and systematic, respectively, for ${ {\mathcal A}_{CP}^{\rm sig}}$; while uncertainties for ${\mathcal B}^+_{\rm sig}$ and ${\mathcal B}^-_{\overline{\rm sig}}$ are only statistical.
}\label{tab:CP}
\resizebox{!}{1.9cm}{
\begin{tabular}{lrrc}
  \hline\hline
\multicolumn{1}{c} {Decay} & \multicolumn{1}{c} {${\mathcal B}^+_{\rm sig}$($\times 10^{-4}$)}  & \multicolumn{1}{c} {${\mathcal B}^-_{\overline{\rm sig}}$($\times 10^{-4}$)}  & ${ {\mathcal A}_{CP}^{\rm sig}}$ (\%) \\  \hline
$\pi^+\pi^-\pi^0$ &$134.8\pm1.8$ & $133.3\pm1.8$ & $+0.6\pm0.9\pm0.4$ \\
$\pi^+\pi^-2\pi^0$ &$97.1\pm2.6$ & $102.3\pm2.7$ & $-2.6\pm1.9\pm0.7$ \\
$2\pi^+\pi^-$ &$33.5\pm1.0$ & $32.7\pm1.0$ & $+1.2\pm2.1\pm0.6$ \\
$\pi^+2\pi^0$ &$48.9\pm1.8$ & $43.4\pm1.7$ & $+6.0\pm2.7\pm0.5$ \\
$2\pi^+\pi^-\pi^0$ &$117.7\pm3.0$ & $116.8\pm3.0$ & $+0.4\pm1.8\pm0.8$ \\
$2\pi^+\pi^-2\pi^0$ &$102.7\pm5.6$ & $111.6\pm5.8$ & $-4.2\pm3.8\pm1.3$ \\
\hline\hline
\end{tabular}
}
\end{table}

To summarize, by analyzing $2.93\,\rm fb^{-1}$ of $e^+e^-$ annihilation data recorded at $\sqrt{s}=3.773$\,GeV with the BESIII detector, we present
the first absolute measurements of the BFs of twenty Cabibbo-suppressed hadronic $D^{0(+)}$ decays involving multiple pions.
For $D^0\to \pi^+\pi^-\pi^0$, $\pi^+\pi^-2\pi^0$, $2\pi^+2\pi^-\pi^0$, and $D^+\to 2\pi^+\pi^-$, $\pi^+2\pi^0$, $2\pi^+\pi^-\pi^0$,
$3\pi^+2\pi^-$, the BF precisions are improved by factors of 1.2-2.9 compared to the world average values based on relative measurements. For the other 13 decay modes, the BFs are measured for the first time.
The reported BFs offer important input for reliable estimations of potential background sources in the precision measurements of $B$ and $D$ decays, especially to properly evaluate the tensions found in the LFU tests with semileptonic $B$ decays.
Amplitude analyses of these multi-body decays with larger data samples available in the near future~\cite{bes3-white-paper,belle2-white-paper}
will open an opportunity to precisely extract more quasi-two-body hadronic $D^{0(+)}$ decay rates, e.g.~$D^+\to \rho^+\pi^0$. Detailed knowledge of these hadronic $D^{0(+)}$ decays is essential to
deeply explore quark U-spin and SU(3)-flavor symmetry breaking effects and thereby improve the predictions of $CP$ violation in the charm sector~\cite{ref5,theory_1,theory_2}. Additionally, the asymmetries of the charge-conjugated BFs of the six $D^{0(+)}$ decays with largest yields are determined.  No statistically significant $CP$ violation is observed.

The BESIII collaboration thanks the staff of BEPCII and the IHEP computing center for their strong support. This work is supported in part by National Key R\&D Program of China under Contracts Nos. 2020YFA0406400 and 2020YFA0406300; National Natural Science Foundation of China (NSFC) under Contracts Nos. 11625523, 11635010, 11735014, 11822506, 11835012, 11935015, 11935016, 11935018, 11961141012, 12022510, 12025502, 12035009, 12035013, 12061131003, 12192260, 12192261, 12192262, 12192263, 12192264, 12192265; the Chinese Academy of Sciences (CAS) Large-Scale Scientific Facility Program; Joint Large-Scale Scientific Facility Funds of the NSFC and CAS under Contracts Nos. U1732263, U1832207, U1932102; CAS Key Research Program of Frontier Sciences under Contract No. QYZDJ-SSW-SLH040; 100 Talents Program of CAS; INPAC and Shanghai Key Laboratory for Particle Physics and Cosmology; ERC under Contract No. 758462; European Union Horizon 2020 research and innovation programme under Contract No. Marie Sklodowska-Curie grant agreement No 894790; German Research Foundation DFG under Contracts Nos. 443159800, Collaborative Research Center CRC 1044, FOR 2359, FOR 2359, GRK 2149; Istituto Nazionale di Fisica Nucleare, Italy; Ministry of Development of Turkey under Contract No. DPT2006K-120470; National Science and Technology fund; Olle Engkvist Foundation under Contract No. 200-0605; STFC (United Kingdom); The Knut and Alice Wallenberg Foundation (Sweden) under Contract No. 2016.0157; The Royal Society, UK under Contracts Nos. DH140054, DH160214; The Swedish Research Council; U. S. Department of Energy under Contracts Nos. DE-FG02-05ER41374, DE-SC-0012069.

\clearpage
\appendix
\onecolumngrid
\section*{Supplemental material}

\begin{figure*}[htp]
  \centering
  \includegraphics[width=0.5\linewidth]{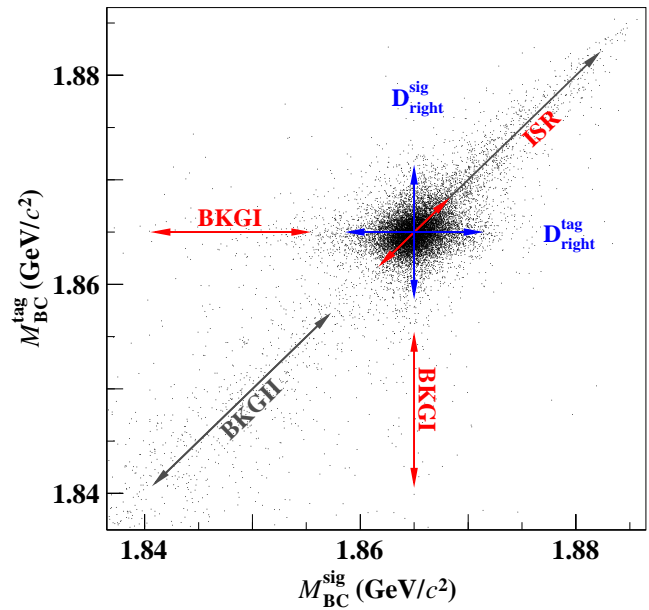}
  \caption{
    The $M_{\rm BC}^{\rm tag}$
    versus $M_{\rm BC}^{\rm sig}$ distribution of the accepted DT candidates of $D^0\to \pi^+\pi^-\pi^0$ in data.
    Here, ISR denotes initial state radiation, which spreads along the diagonal direction and extends to the higher $M_{\rm BC}$ sides. The $D^{\rm sig}_{\rm right}$ and $D^{\rm tag}_{\rm right}$ denotes the signal spreading around $M_{\rm BC}^{\rm sig} = M_{D}$ and $M_{\rm BC}^{\rm tag} = M_{D}$.
}
\label{fig:2D_MBC}
\end{figure*}

Figure~\ref{fig:2D_MBC} shows the $M_{\rm BC}^{\rm tag}$
versus $M_{\rm BC}^{\rm sig}$ distribution of the accepted DT candidates in data. Detailed selection criteria can be found in context of the main paper.

\begin{figure*}[htp]
  \centering
  \includegraphics[width=1.0\linewidth]{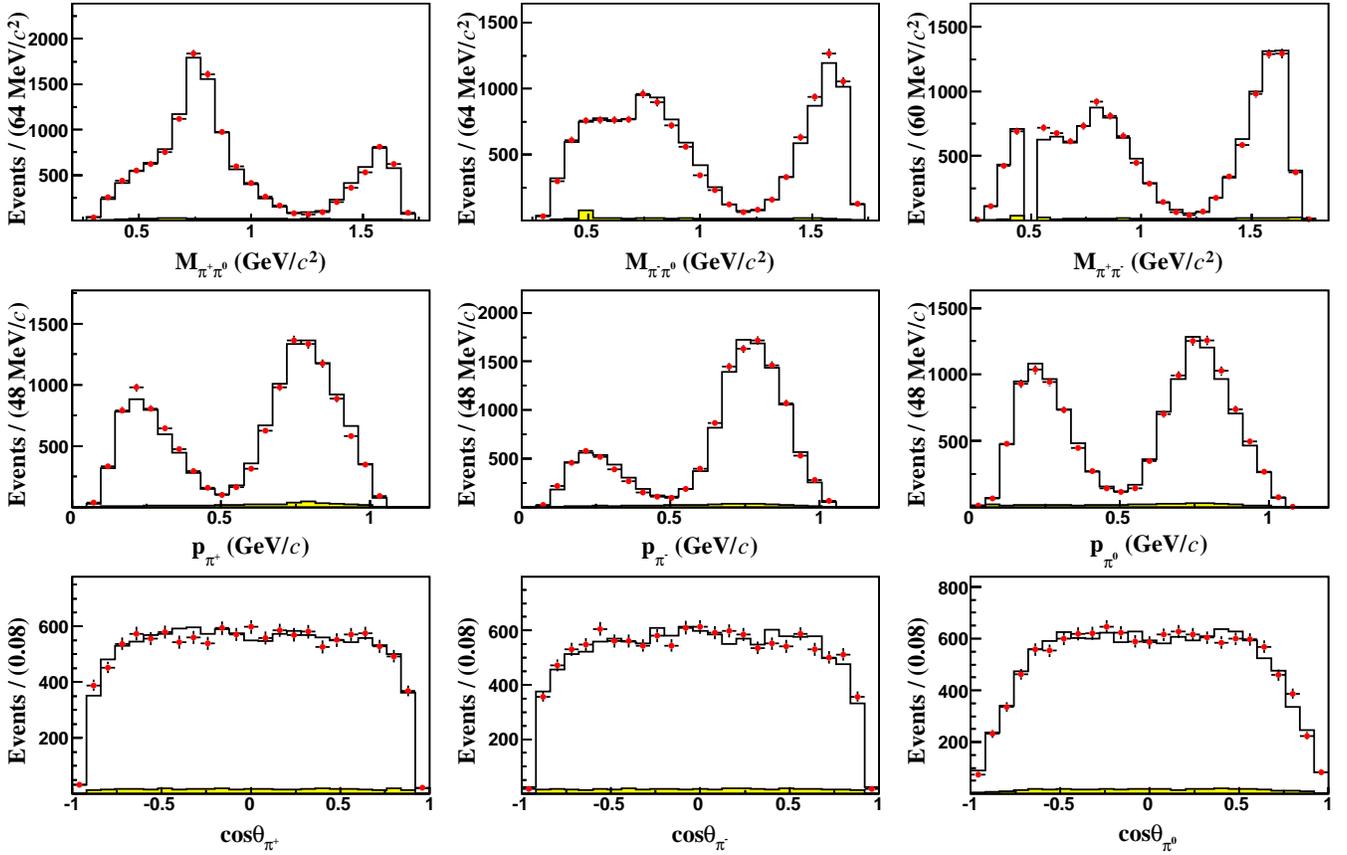}
  \caption{Comparisons of the distributions of invariant masses of two-body particle combinations, momenta and $\cos\theta$ of daughter particles for the $D^0\to \pi^+\pi^-\pi^0$ candidates between data (points with error bars) and the BODY3 signal MC events (black solid line histograms) plus the MC-simulated backgrounds from the inclusive MC sample (yellow filled histograms). }
\label{fig:pipipi0}
\end{figure*}

\begin{figure*}[htp]
  \centering
  \includegraphics[width=1.0\linewidth]{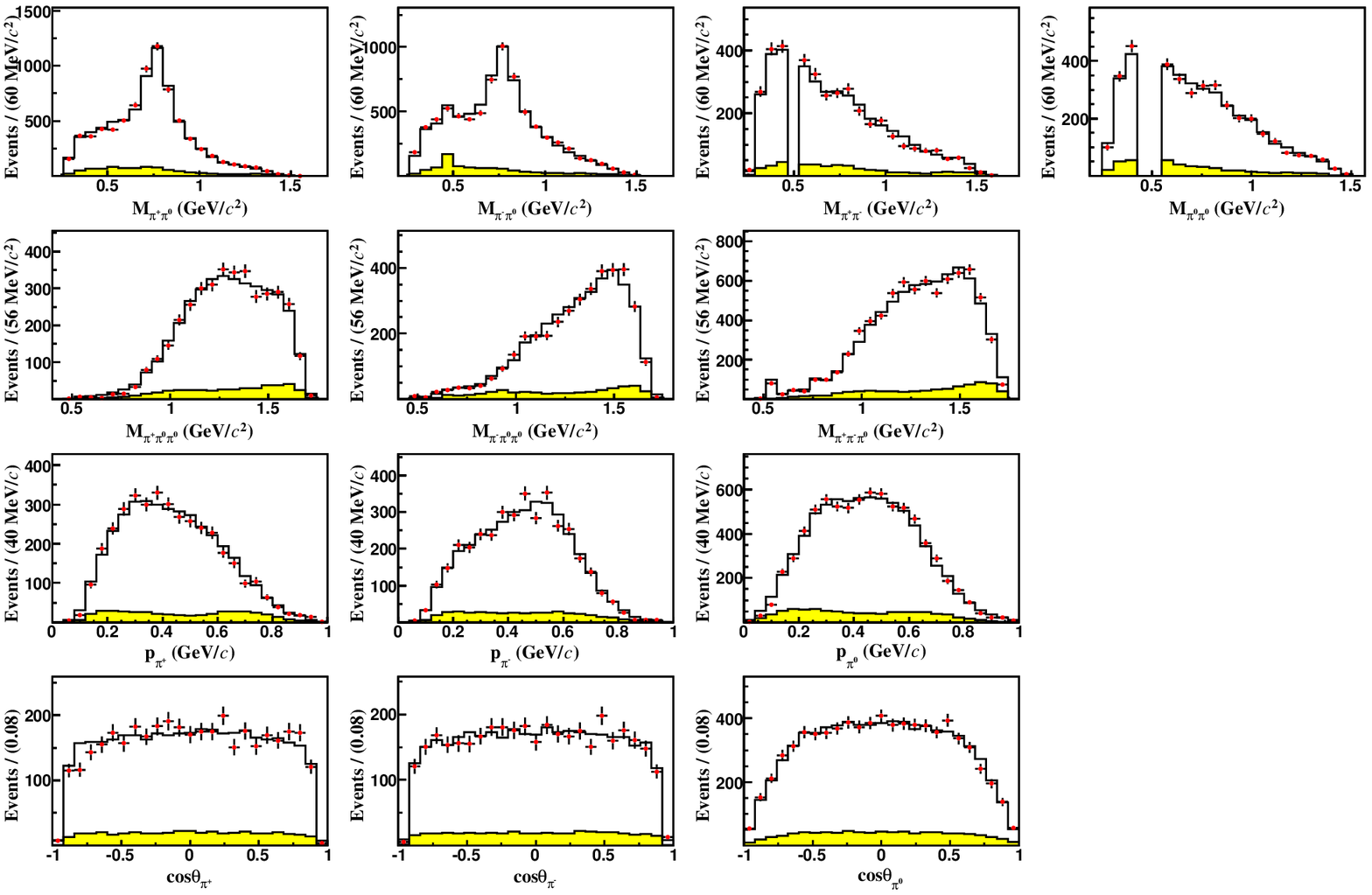}
  \caption{Comparisons of the distributions of invariant masses of two- or three-body particle combinations, momenta and $\cos\theta$ of daughter particles for the $D^0\to \pi^+\pi^-2\pi^0$ candidates between data (points with error bars) and the mixing signal MC events (black solid line histograms) plus the MC-simulated backgrounds from the inclusive MC sample (yellow filled histograms). }
\label{fig:pipipi0pi0}
\end{figure*}

\begin{figure*}[htp]
  \centering
  \includegraphics[width=1.0\linewidth]{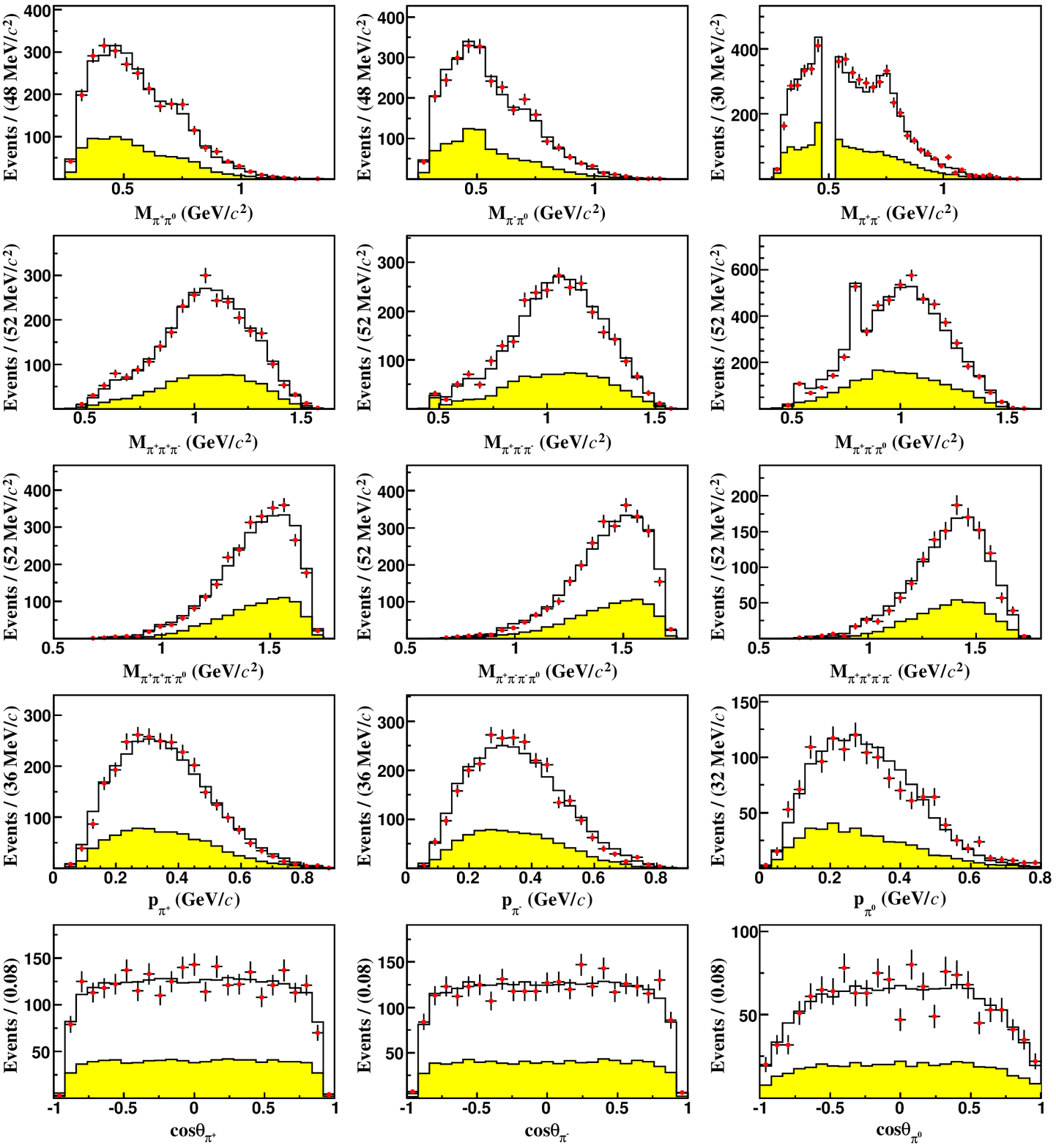}
  \caption{Comparisons of the distributions of invariant masses of two-, three-, or four-body particle combinations, momenta and $\cos\theta$ of daughter particles for the $D^0\to 2\pi^+2\pi^-\pi^0$ candidates between data (points with error bars) and the mixing signal MC events (black solid line histograms) plus the MC-simulated backgrounds from the inclusive MC sample (yellow filled histograms). }
\label{fig:pipipipipi0}
\end{figure*}

\begin{figure*}[htp]
  \centering
  \includegraphics[width=1.0\linewidth]{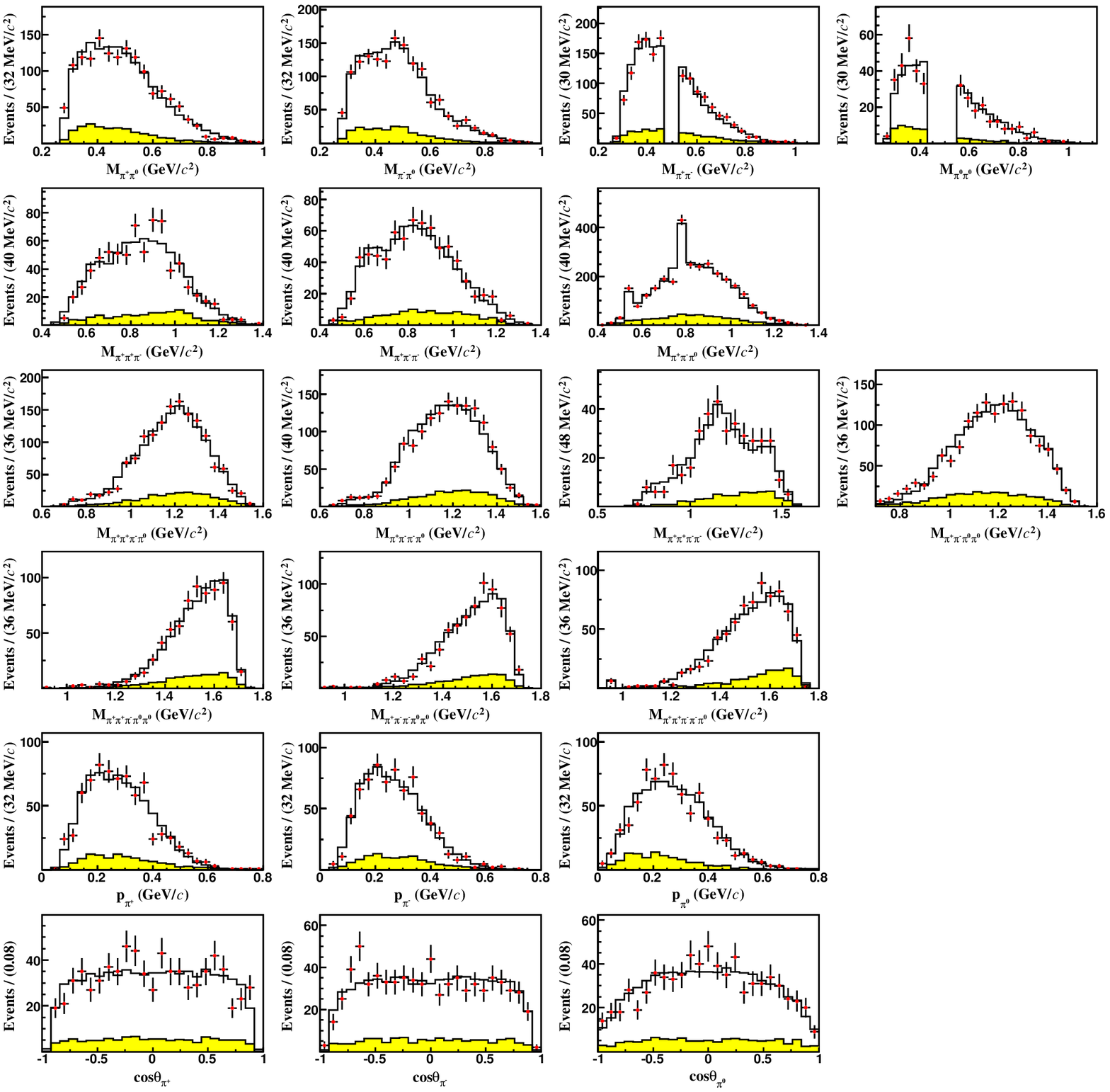}
  \caption{Comparisons of the distributions of invariant masses of two-, three-, four- or five-body particle combinations, momenta and $\cos\theta$ of daughter particles for the $D^0\to 2\pi^+2\pi^-2\pi^0$ candidates between data (points with error bars) and the mixing signal MC events (black solid line histograms) plus the MC-simulated backgrounds from the inclusive MC sample (yellow filled histograms). }
\label{fig:pipipipipi0pi0}
\end{figure*}

\begin{figure*}[htp]
  \centering
  \includegraphics[width=1.0\linewidth]{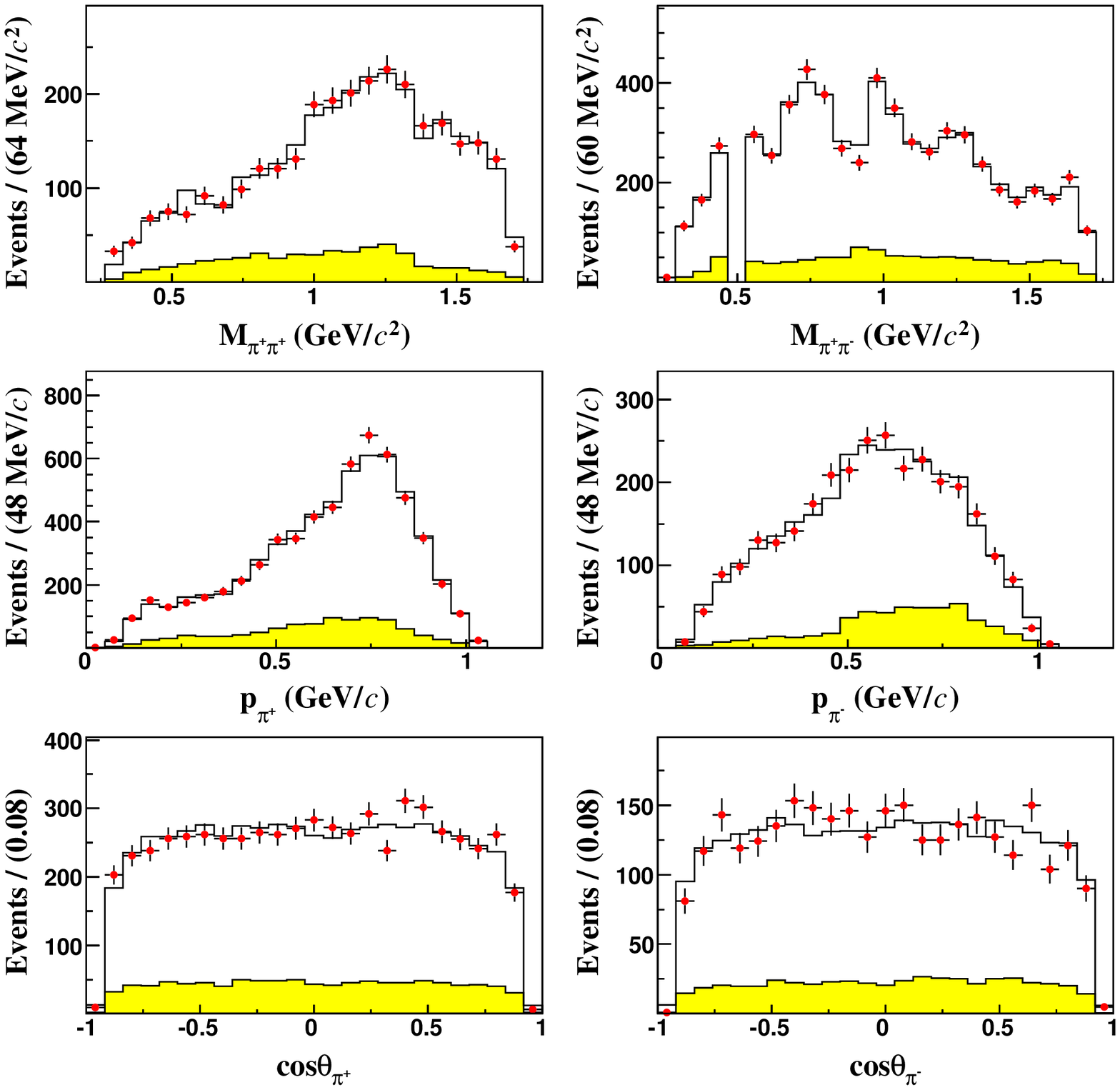}
  \caption{Comparisons of the distributions of invariant masses of two-body particle combinations, momenta and $\cos\theta$ of daughter particles for the $D^+\to 2\pi^+\pi^-$ candidates between data (points with error bars) and the BODY3 signal MC events (black solid line histograms) plus the MC-simulated backgrounds from the inclusive MC sample (yellow filled histograms). }
\label{fig:pipipi}
\end{figure*}

\begin{figure*}[htp]
  \centering
  \includegraphics[width=1.0\linewidth]{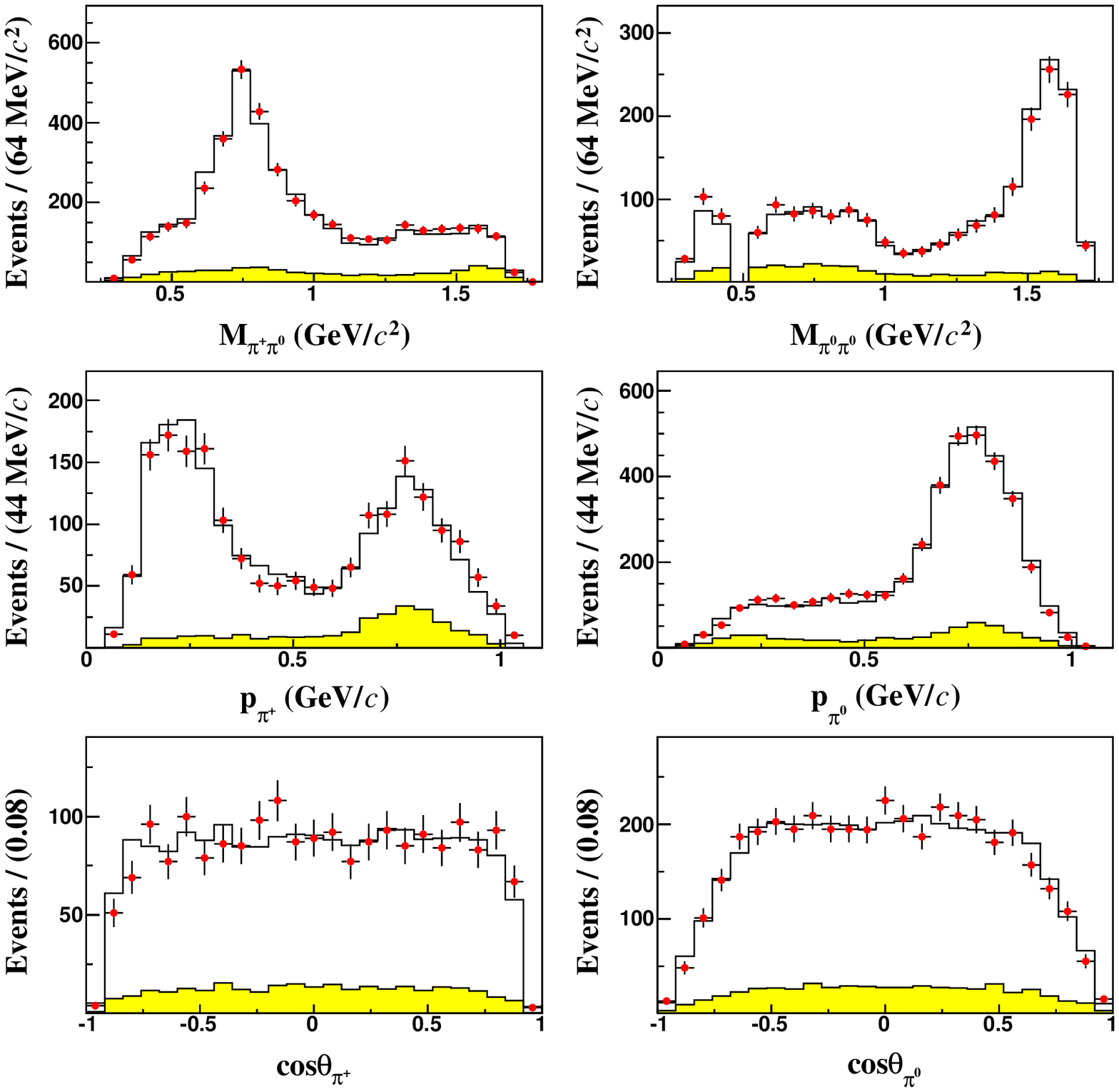}
  \caption{Comparisons of the distributions of invariant masses of two-body particle combinations, momenta and $\cos\theta$ of daughter particles for the $D^+\to \pi^+2\pi^0$ candidates between data (points with error bars) and the BODY3 signal MC events (black solid line histograms) plus the MC-simulated backgrounds from the inclusive MC sample (yellow filled histograms). }
\label{fig:pipi0pi0}
\end{figure*}

\begin{figure*}[htp]
  \centering
  \includegraphics[width=1.0\linewidth]{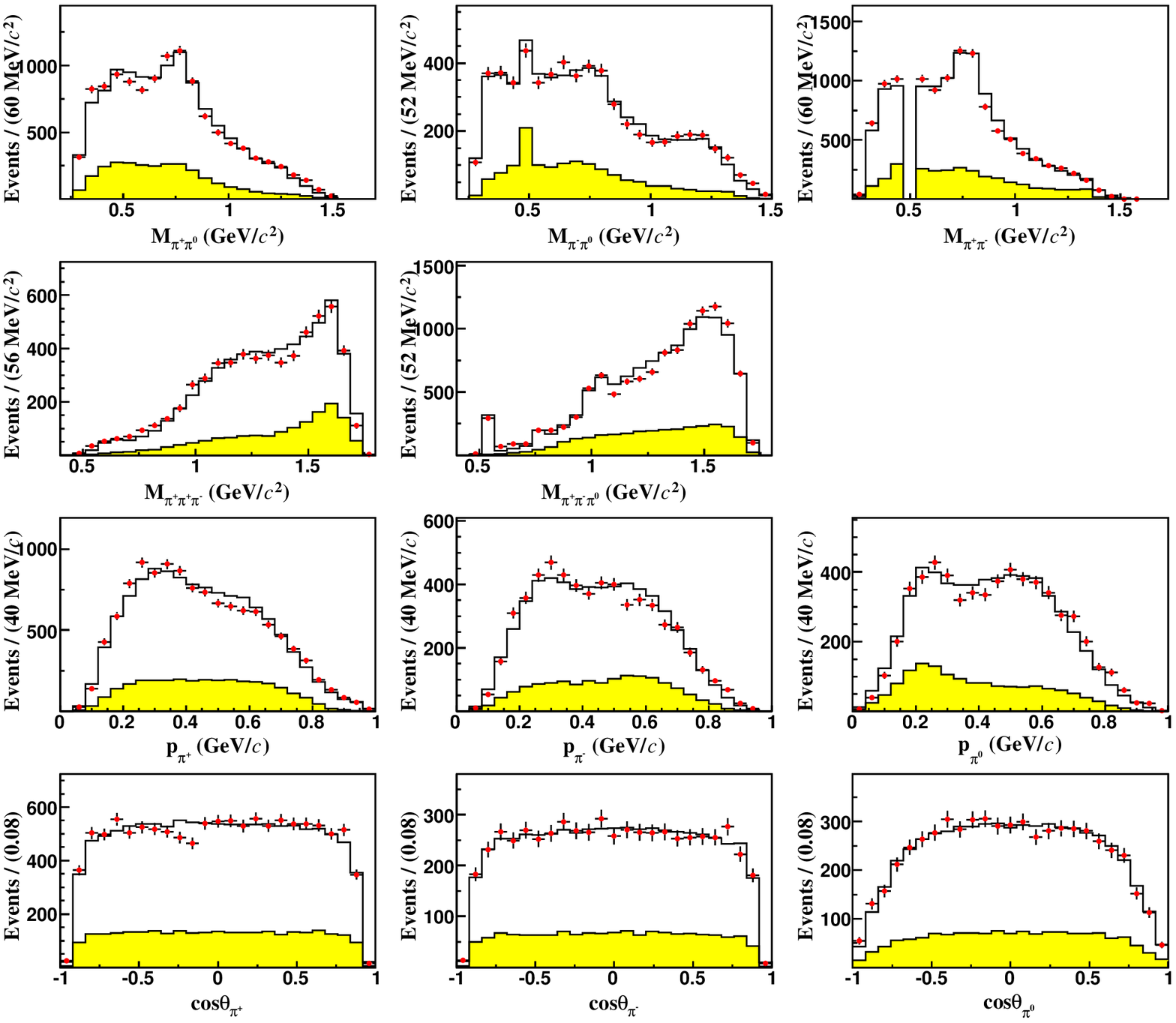}
  \caption{Comparisons of the distributions of invariant masses of two- or three-body particle combinations, momenta and $\cos\theta$ of daughter particles for the $D^+\to 2\pi^+\pi^-\pi^0$ candidates between data (points with error bars) and the mixing signal MC events (black solid line histograms) plus the MC-simulated backgrounds from the inclusive MC sample (yellow filled histograms). }
\label{fig:pipipipi0}
\end{figure*}

\begin{figure*}[htp]
  \centering
  \includegraphics[width=1.0\linewidth]{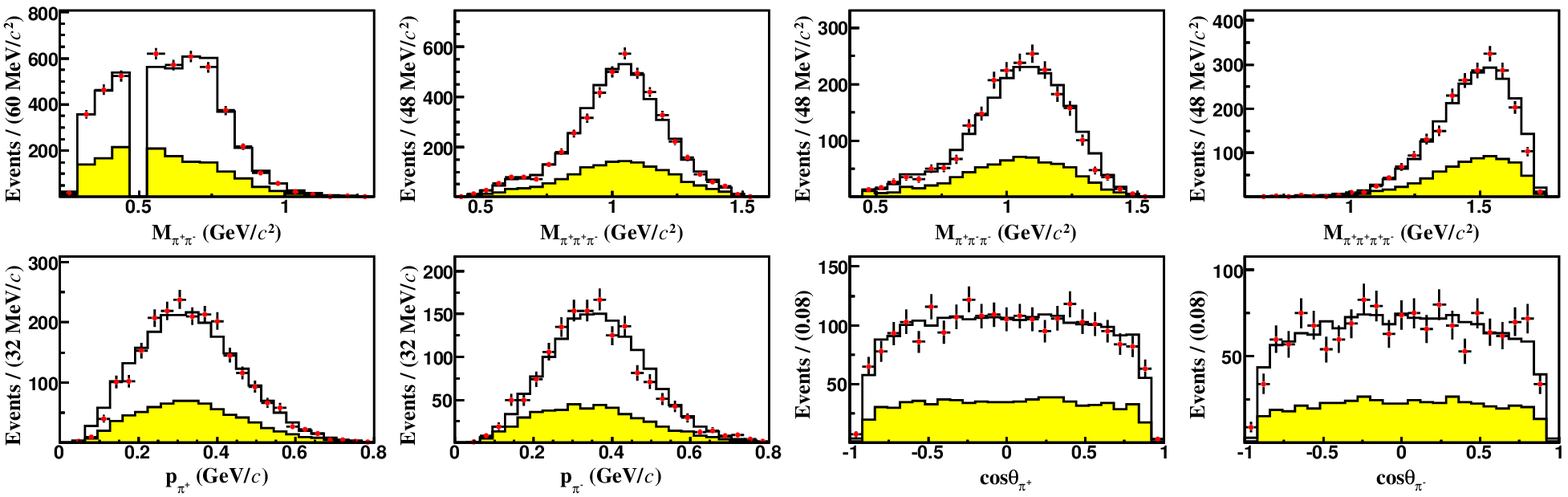}
  \caption{Comparisons of the distributions of invariant masses of two- ,three- or four-body particle combinations, momenta and $\cos\theta$ of daughter particles for the $D^+\to 3\pi^+2\pi^-$ candidates between data (points with error bars) and the mixing signal MC events (black solid line histograms) plus the MC-simulated backgrounds from the inclusive MC sample (yellow filled histograms). }
\label{fig:pipipipipi}
\end{figure*}

\begin{figure*}[htp]
  \centering
  \includegraphics[width=0.95\linewidth]{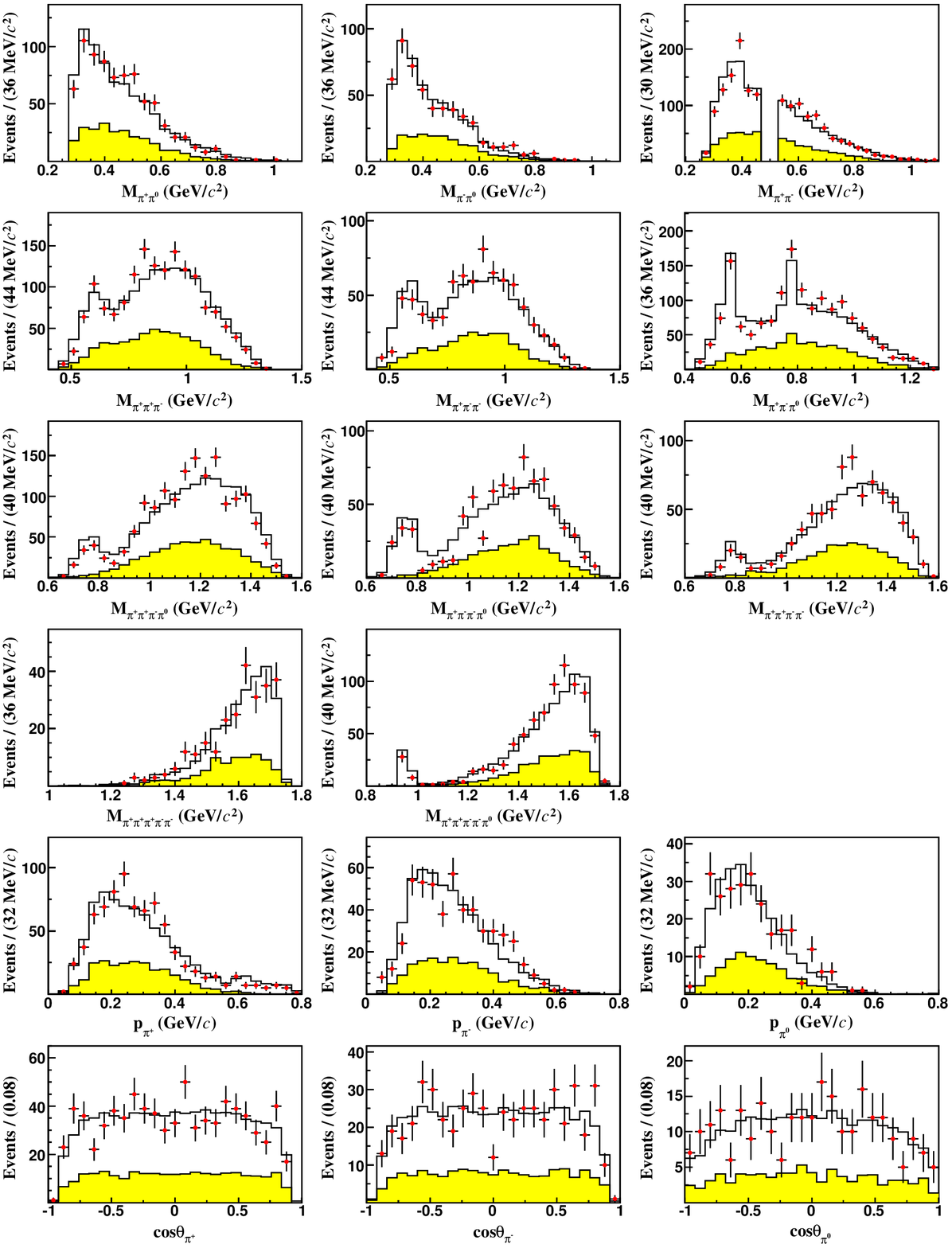}
  \caption{Comparisons of the distributions of invariant masses of two-, three-, four- or five-body particle combinations, momenta and $\cos\theta$ of daughter particles for the $D^+\to 3\pi^+2\pi^-\pi^0$ candidates between data (points with error bars) and the mixing signal MC events (black solid line histograms) plus the MC-simulated backgrounds from the inclusive MC sample (yellow filled histograms). }
\label{fig:pipipipipipi0}
\end{figure*}

Figures \ref{fig:pipipi0} to \ref{fig:pipipipipipi0} show comparisons between data and MC simulations for the distributions of invariant mass spectra of two-, three-, four- or five-body particle combinations, momenta and $\cos\theta$ of daughter particles for the signal DT candidates with more than 100 signal events.  The candidates must satisfy additional requirements of $|M^{\rm tag(sig)}_{\rm BC}-M_D|<0.006$ GeV/$c^2$ and multiple possible combinations of daughter particles are all plotted when relevant (e.g., all four $\pi^+\pi^-$ combinations for $D^0 \to 2\pi^+2\pi^-\pi^0$, etc.)

Table~\ref{CP_pam} summarizes
the ST yields of $CP\pm$ tags from the fits to the $M^{\rm tag}_{\rm BC}$ distributions of the accepted ST candidates,
the DT yields tagged by $CP\pm$ tags from the 2D fits to the $M^{\rm tag}_{\rm BC}$ versus $M^{\rm sig}_{\rm BC}$ distributions of the accepted DT candidates, and  the quantum correlation~(QC) factors obtained with the
same method as described in Ref.~\cite{QC-factor} and the necessary parameters quoted from Refs.~\cite{R-ref1,R-ref2,R-ref3}.

Table~\ref{tab:significance} summarizes the statistical significances of the decay modes.

Table~\ref{tab:sys} summarizes the systematic uncertainties for various sources in the measurements of BFs, which
are assigned relative to the measured BFs.
They are from the ST yield ($N_{\rm tag}$), $\pi^\pm$ tracking efficiency, $\pi^\pm$ PID efficiency, $\pi^0$ and $\eta$ reconstruction efficiency, daughter BFs, $\Delta E$ requirement, $K^0_S$ rejection, MC statistics, MC generator, 2D fit, and strong phase.
For each signal decay, the total uncertainty is obtained by quadratically adding all uncertainties.

\begin{table}[htbp]
\centering
\caption{
Summary of the ST yields of $CP\mp$ tags ($S^\pm_{\rm measured}$),
the DT yields tagged by $CP\mp$ tags ($M^\pm_{\rm measured}$),
the $CP+$ fraction ($f_{CP+}$),
and the QC factor ($f_{\rm QC}$).
The uncertainties are statistical only.
A ``/'' denotes unmeasured quantites, occuring for one mode with
a high-precision extrernal result and for the two $CP$-eigenstates.}
\label{CP_pam}
\centering
\scalebox{1.05}{
\begin{tabular}{|c|c|c|c|c|c|}
\hline
\multirow{4}*{$CP$ mode}&$\bar{D}^0\to K^+K^-$ ($CP+$)&$\bar{D}^0\to K^0_S\pi^0$ ($CP-$)
&\multicolumn{3}{|c|}{}\\
\cline{2-3}
\multicolumn{1}{|c|}{}                  &$S^{-}_{\rm measured}$          &$S^{+}_{\rm measured}$&\multicolumn{3}{|c|}{}\\ \cline{2-3}
\multicolumn{1}{|c|}{}                  &57779$\pm$287    &70512$\pm$311&\multicolumn{3}{|c|}{} \\
\hline
 \hline
\multicolumn{1}{|c|}{Decay mode}&$M^{-}_{\rm measured}$&$M^{+}_{\rm measured}$& $f_{CP+}$ & $f_{\rm QC}$ (\%) & Uncertainty (\%) \\ \hline
$D^0\to \pi^+\pi^-\pi^0$    &/       &/      & $0.973\pm0.017$~\cite{CP-3pi}                              &$ 93.5\pm0.5$&  0.5      \\ \hline
$D^0\to \pi^+\pi^-2\pi^0$   &$65.7\pm11.1$       &$169.8\pm13.9$      & $0.682\pm0.077$                   &$ 97.4\pm0.7$&  0.7      \\ \hline
$D^0\to 4\pi^0$             &/       &/      & $1$                                                        &$ 93.1\pm0.5$&  0.5      \\ \hline
$D^0\to 3\pi^0\eta$         &/       &/      & $1$                                                        &$ 93.1\pm0.5$&  0.5      \\ \hline
$D^0\to 2\pi^+2\pi^-\pi^0$  &$37.8\pm8.3$    &$35.5\pm6.6$   & $0.438\pm0.104$                            &$100.9\pm0.9$&  0.9      \\ \hline
$D^0\to \pi^+\pi^-3\pi^0$   &$5.2^{+3.5}_{-2.8}$       &$6.8^{+3.4}_{-2.7}$   & $0.520^{+0.338}_{-0.269}$ &$ 99.7^{+3.0}_{-2.4}$&  3.0      \\ \hline
$D^0\to 2\pi^+2\pi^-2\pi^0$ &$3.5^{+2.8}_{-2.1}$       &$15.9\pm3.7$          & $0.790^{+0.269}_{-0.255}$ &$ 95.9^{+2.2}_{-2.1}$&  2.2      \\ \hline
\end{tabular}
}
\end{table}

\begin{table}[http]
\centering
\caption{Statistical significances $(\sigma)$ for various decay modes.}
\label{tab:significance}
\centering
\scalebox{0.98}{
\begin{tabular}{|c|c|c|c|c|c|c|c|c|c|c|}
\hline
Decay mode
&$\pi^+\pi^-\pi^0$
&$\pi^+\pi^-2\pi^0$
&$\pi^+\pi^-2\eta$
&$4\pi^0$
&$3\pi^0\eta$
&$2\pi^+2\pi^-\pi^0$
&$2\pi^+2\pi^-\eta$
&$\pi^+\pi^-3\pi^0$
&$2\pi^+2\pi^-2\pi^0$
&$2\pi^+\pi^-$  \\
\hline
Significance     &$>10$ &$>10$ &8.7 &$>10$ &$>10$ &$>10$ &9.1 &$>10$ &$>10$ &$>10$   \\
\hline
\hline
Decay mode
&$\pi^+2\pi^0$
&$2\pi^+\pi^-\pi^0$
&$\pi^+3\pi^0$
&$3\pi^+2\pi^-$
&$2\pi^+\pi^-2\pi^0$
&$2\pi^+\pi^-\pi^0\eta$
&$\pi^+4\pi^0$
&$\pi^+3\pi^0\eta$
&$3\pi^+2\pi^-\pi^0$
&$2\pi^+\pi^-3\pi^0$   \\
\hline
Significance     &$>10$ &$>10$ &$>10$ &$>10$ &$>10$ &$>10$ &6.8 &9.9 &$>10$ &$>10$   \\
\hline

\end{tabular}
}
\end{table}

\begin{table}[http]
\centering
\caption{
Systematic uncertainties (\%) in the measurements of the
BFs for various decay modes.
}
\label{tab:sys}
\centering
\scalebox{0.92}{
\begin{tabular}{|c|c|c|c|c|c|c|c|c|c|c|}
\hline
\multirow{2}{*}{Source}&\multicolumn{9}{c}{$D^0\to$}&\multicolumn{1}{|c|}{$D^+\to$}\\ \cline{2-11}
 &$\pi^+\pi^-\pi^0$&$\pi^+\pi^-2\pi^0$&
$\pi^+\pi^-2\eta$&
$4\pi^0$&
$3\pi^0\eta$&
$2\pi^+2\pi^-\pi^0$&
$2\pi^+2\pi^-\eta$&
$\pi^+\pi^-3\pi^0$&
$2\pi^+2\pi^-2\pi^0$&$2\pi^+\pi^-$\\
\hline
$N_{\rm tag}$                                  &0.5 &0.5 &0.5 &0.5 &0.5 &0.5 &0.5 &0.5 &0.5 &0.5 \\
$\pi^\pm$ tracking                             &0.4 &0.4 &0.7 &--  &--  &0.8 &1.0 &0.4 &0.8 &0.6 \\
$\pi^\pm$ PID                                  &0.4 &0.4 &0.4 &--  &--  &0.8 &0.8 &0.4 &0.8 &0.6 \\
$\pi^0/\eta$ reconstruction                    &0.4 &1.3 &1.6 &2.3 &2.5 &0.6 &0.7 &1.9 &1.3 &-- \\
2D fit                                         &0.3 &0.6 &4.4 &3.4 &4.7 &1.7 &4.0 &3.5 &2.3 &0.3 \\
$\Delta E^{\rm sig}$ cut                       &0.1 &0.1 &1.3 &0.1 &0.1 &0.2 &0.8 &0.1 &0.1 &0.3 \\
MC generator                                   &0.6 &1.5 &0.2 &3.4 &2.8 &1.1 &1.5 &5.3 &1.4 &0.8 \\
$K_{S}^{0}$ rejection                          &0.1 &0.4 &--  &7.4 &3.5 &3.2 &8.1 &4.7 &1.6 &0.5 \\
Strong phase                                   &0.5 &0.7 &--  &0.5 &0.5 &0.9 &--  &3.0 &2.2 &--  \\
MC statistics                                  &0.2 &0.4 &0.5 &0.8 &0.7 &0.5 &0.6 &0.8 &1.0 &0.2 \\Daughter $\mathcal B$                          &0.03&0.07&1.02&0.14&0.52&0.03&0.51&0.10&0.07&-- \\
\hline
Total                                          &1.2 &2.4 &5.1 &9.2 &7.0 &4.2 &9.4 &8.7 &4.3 &1.4 \\
\hline
\multirow{2}{*}{Source}&\multicolumn{10}{c|}{$D^+\to$}\\ \cline{2-11}
&
$\pi^+2\pi^0$&
$2\pi^+\pi^-\pi^0$&
$\pi^+3\pi^0$&
$3\pi^+2\pi^-$&
$2\pi^+\pi^-2\pi^0$&
$2\pi^+\pi^-\pi^0\eta$&
$\pi^+4\pi^0$&
$\pi^+3\pi^0\eta$&
$3\pi^+2\pi^-\pi^0$&
$2\pi^+\pi^-3\pi^0$\\ \hline
$N_{\rm tag}$                                  &0.5 &0.5 &0.5 &0.5 &0.5 &0.5 &0.5 &0.5 &0.5 &0.5 \\
$\pi^\pm$ tracking                             &0.2 &0.6 &0.2 &1.1 &0.6 &0.8 &0.2 &0.2 &1.2 &0.6 \\
$\pi^\pm$ PID                                  &0.2 &0.6 &0.2 &1.0 &0.6 &0.6 &0.2 &0.2 &1.0 &0.6 \\
$\pi^0/\eta$ reconstruction                    &1.0 &0.6 &1.8 &--  &1.4 &1.4 &2.7 &3.4 &0.8 &2.5 \\
2D fit                                         &0.3 &0.6 &1.7 &2.1 &1.8 &2.3 &3.1 &3.4 &2.8 &2.7 \\
$\Delta E^{\rm sig}$ cut                       &0.1 &0.1 &0.1 &0.2 &0.1 &0.5 &0.1 &0.1 &0.4 &0.1 \\
MC generator                                   &1.5 &0.8 &1.5 &3.0 &0.7 &0.3 &4.2 &5.7 &0.5 &1.7 \\
$K_{S}^{0}$ rejection                          &--  &0.8 &1.1 &3.4 &0.8 &--  &10.2&--  &5.6 &2.2 \\
Strong phase                                   &--  &--  &--  &--  &--  &--  &--  &--  &--  &--  \\
MC statistics                                  &0.3 &0.3 &0.6 &0.4 &0.6 &0.6 &1.3 &0.8 &0.8 &0.9 \\
Daughter $\mathcal B$                          &0.07&0.03&0.10&--  &0.07&0.51&0.14&0.52&0.03&0.10\\
\hline
Total                                          &1.9 &1.8 &3.2 &5.3 &2.8 &3.1 &11.9&7.5 &6.6 &4.8 \\
\hline
\end{tabular}
}
\end{table}

\end{document}